# Broadly tunable (440-670 nm) solid-state organic laser with disposable capsules

Oussama Mhibik[1,2], Tatiana Leang[1,2], Alain Siove[1,2], Sébastien Forget[a) 1,2] and Sébastien Chénais[1,2]

[1] Université Paris 13, Sorbonne Paris Cité, Laboratoire de Physique des Lasers, F-93430, Villetaneuse, France

[2] CNRS, UMR 7538, LPL, F-93430, Villetaneuse, France

**Abstract**

**An innovative concept of thin-film organic solid-state laser is proposed, with diffraction-limited output and a broad tuning range covering the visible spectrum under UV optical pumping. The laser beam is tunable over 230 nm, from 440 to 670 nm, with a 3 nm (FWHM) typical spectral width. The structure consists of a compact fixed bulk optical cavity, a polymeric intracavity etalon for wavelength tuning, as well as five different disposable glass slides coated with a dye-doped polymer film, forming a very simple and low-cost gain medium. The use of interchangeable/disposable "gain capsules" is an alternative solution to photodegradation issues, since gain chips can be replaced without**

**realignment of the cavity. The laser lifetime of a single chip in ambient conditions and without encapsulation was extrapolated to be around $10^7$ pulses at a microjoule energy-per-pulse level.**

Widely tunable laser sources in the visible part of the spectrum are required for many applications such as, for instance, spectroscopy[1-3] or bio-chemo sensing[4-6]. While tunable liquid dye lasers have been reigning for decades over this spectral region in virtue of their unrivalled spectral coverage and wavelength agility, they require complex and bulky flow systems and are plagued by stability and toxicity issues. Optical parametric oscillators (OPO) or Generators (OPG)[7,8] provide a solid-state alternative solution to get tunable visible coherent radiation, but despite many recent improvements[9], they still remain complex, cumbersome and expensive solutions. The landscape of tunable visible sources seemed to change deeply with the advent of supercontinuum sources, which offer broadband spatially-coherent light covering the whole visible and IR spectrum (400-1800 nm); however these sources have still low output powers per unit nm (typ. ~mW/nm), and are hence not ideal whenever only a specific spectral region has to be addressed, since in that case only a tiny fraction of the light is useful after filtering. Tunable visible sources based on nonlinear fibers and frequency conversion recently appeared on the marketplace but are available at a cost comparable with that of dye lasers or OPOs. In pulsed regime, organic solid-state lasers (OSSLs)[10,11] are then very appealing devices which combine the advantages of dye lasers with those of conventional solid-state lasers, with a "low-cost" advantage. Two types of OSSLs can be distinguished: lasers with bulk (≈ cm) dye-doped polymer chips, and thin-film based lasers (which include organic semiconductor lasers[11]). Bulk OSSLs are characterized by an open resonator, allowing for a perfect beam quality and enabling an easy insertion of optical elements inside the cavity for wavelength tuning[12]. These lasers can attain nearly four-level limit efficiencies[13], however the gain chip (typically a dye-doped polymer block prepared by radical polymerization or sol-gel[14]) might be long and complex to produce, and mandatory requires optical polishing before use[15]. Conversely, the archetypal example of a thin-film-based OSSL, *i.e.* an organic semiconductor distributed feedback (DFB) laser, will have a low-loss resonator with a very low laser threshold, but will not be so easily tuned. Continuous wavelength tunability of a solid-state organic DFB laser requires modifying the active medium shape in some way, either through a modification of the effective index of the layer (which can be done by playing on the index[16,17], the film thickness[18,19] or the grating depth[20]), or by mechanical means upon applying a stress to the active layer in order to induce a shift of the Bragg wavelength[21,22]. Their tuning range is limited to a few nanometers for a single grating, and at most limited by the gain bandwidth of a given material; a full coverage of the visible spectrum requires dozens of gratings with different pitches, leading to increased cost and fabrication issues. Moreover, DFB laser devices have a few drawbacks like a poorly-defined spatial beam quality[23,24], a low optical conversion efficiency and a related low output energy.

In this letter, we report on the realization of a simple, diffraction-limited, efficient and broadly tunable organic solid state laser covering the whole visible spectrum. We use an innovative and cost-effective approach to bypass the photodegradation issue which is common to all organic lasers, based on the use of a very simple "gain capsule" that can be produced at negligible cost, and which can be easily replaced or exchanged without alignment once the material is degraded. Instead of waging a hopeless battle against the inevitable limited lifetime of any organic material under intense optical pumping, the use of such a "disposable" gain element takes full benefit of the low cost of organic materials.

Our experimental setup is described in Fig.1. It consists of a modified version of the vertical external-cavity surface-emitting organic laser (VECSOL) architecture[26] recently demonstrated in our group. It shares the same general properties as vertical external-cavity surface-emitting lasers (VECSELs): diffraction-limited output, high efficiency, power scaling capability upon a simple increase of pump spot size (without higher photodegradation rate), possibility of long (cm) cavities that enable the insertion of intracavity elements, such as spatial or spectral filters. However, as antireflection coatings are not practically implemented onto low-index organic thin films, the VECSOL emission spectrum is comb-shaped since the 17-μm-thick active layer, directly deposited onto the rear mirror, acts as a Fabry-Perot intracavity etalon. In order to ensure a continuous tunability with a reduced linewidth the original structure was then slightly modified: the gain medium was physically separated from the rest of the cavity backbone to take the form of a "capsule". This capsule in its simplest form, consists in a glass substrate onto which a film of dye-doped polymer was spin-coated. In order to demonstrate the low-cost and disposable potential, as well as the robustness, of our approach, the results of this paper were obtained with commercially available microscope plain-glass slides as substrates, which cost less than 10 cents each. Lasing was also successively observed with poly(ethylene terephthalate)) (PET) plastic sheets taken from commercial transparencies, coated with the same dye-doped polymer film. The pump laser source is a frequency-tripled Q-switched diode-pumped Nd:YAG laser (Harrier from Quantronix Inc.) providing 355-nm UV radiation with a 20-ns pulsewidth and a 10 Hz repetition rate. The pump beam was focused to a 200-µm-in-diameter waist onto the active "capsule" placed close to the input mirror and positioned at Brewster angle to remove the remaining etalon effect which arises from the index mismatch between the dye-doped polymer film and the glass plate.

The resonator was composed of a highly reflective plane dielectric mirror (R=99.5% in the visible) and a 200-mm radius-of-curvature broadband output coupler (R=98% +/-1% between 400 and 680 nm). The input plane mirror was highly transparent for the pump wavelength (T>95% at 355 nm).

To cover a wavelength range going from blue to red, five dyes belonging to three dye families, namely coumarins, pyrromethenes, and xanthenes have been chosen for their well-established laser performance. Coumarin dyes (coumarin 460 and coumarin 540 A) emit in the blue-green region, pyrromethenes (pyrromethene 567 and pyrromethene 597) emit in yellow-orange region while the chosen xanthene dye (rhodamine 640) emits in the red region. All active materials were dispersed into polymethylmethacrylate (PMMA, molar mass 950000 in anisole solution, from Microchem Inc.) either alone or in the form of donor/acceptor mixtures according to their absorption properties at 355 nm. Depending on the nature of the dye, excitation occurs on the first excited singlet state $S_1$ (the case of coumarin dyes) or on higher excited singlet states $S_{n,\ n>1}$ (n=2 for pyrromethene and xanthene dyes) Materials exhibiting a high absorption cross section at 355 nm were used as delivered, and dispersed at x% into PMMA, x being determined in order to obtain almost complete absorption over a 20-µm thick film. Hence, Coumarin 460 (x=1) and Coumarin 540A (x=2) were used to cover the blue (440-470 nm) and the green (470-550 nm) emission ranges, respectively (pumping on the $S_1$ level). In order to obtain yellow-orange laser emission, pyrromethene 567 (560-580 nm) and pyrromethene 597 (580-620 nm) were used, both of them with x=1, although for these materials pumping takes place on the $S_0 \rightarrow S_2$ transition. In order to address the red part of the spectrum, efficient direct pumping at 355 nm is unfeasible as most materials have too low absorption cross sections. To solve this problem, a donor/acceptor mixture was used: the "donor" dye absorbs the pump light up to its $S_1$ level and transfers its excitation to the "acceptor" dye by a Förster resonant energy transfer (FRET) mechanism[27]. Efficient FRET in a blend system fundamentally requires a good spectral overlap between the emission of the donor and the absorption of the acceptor as well as a good mixing of the two species. For this purpose we used Coumarin 540 A / Rhodamine 640 as a donor/acceptor pair. We blended coumarin 540 A and rhodamine 640 with weight ratios of 1% and 1.9% in PMMA, respectively, a mixture that was experimentally found to guarantee a complete energy transfer.

As discussed in [ref. 28], oscillation buildup considerations lead to laser performances that drop significantly with the cavity length because dye emitters have a short excited state lifetime which is also comparable, in a VECSOL, with the photon cavity lifetime and the pump pulse duration. Hence, the cavity length must be carefully chosen as a compromise between keeping some space to set intracavity elements and maintaining a correct efficiency. Here we chose a pumping source delivering "long" pulses (20 ns FWHM), which relaxes the constraint on cavity length and allows lasing for up to 120-mm-long cavities, *i.e.* more than ten times the distance needed to insert a Brewster-angled gain capsule together with a tuning filter.

Each dye material has typically a bandwidth of 50 nm. To finely tune the wavelength within this range, an etalon with a free spectral range higher than 50 nm is required, which corresponds to a ~2-μm thick etalon assuming an index of 1.5. Such an etalon is difficult to realize from bulk glass plates which can hardly be made thinner than ~20 μm. To overcome this limitation we used a home-made free-standing PMMA film as an etalon: a mixture (1:1) of two PMMA solutions with different viscosities (PMMA A15, molar mass 950000 and PMMA A6 with molar mass 495000, from MicroChem Corp.) were used to obtain homogenous spin-coated films with a thickness in the desired micron range. After spin coating (3500 rpm for 75 s) the PMMA-coated glass samples were transferred to the oven and annealed at 50°C during 30 seconds to evaporate the solvent while keeping the film soft enough to enable the layer to be peeled off gently. The free-standing film was then glued to an annular mount. The thickness was controlled at the center of the mount by measuring the free spectral range in the arm of a spectrophotometer.

Figure 2 shows the lasing spectra recorded using a spectrometer (Jobin Yvon SPEX 270M) with a resolution of 0.8 nm. The wavelength can be continuously tuned over 40 nm for each dye material by tilting the etalon. Each peak has a full width at half maximum (FWHM) of 3 nm due to the weak finesse of the etalon (F≈ 4) and contains many modes of the external cavity, which cannot be resolved here. Note that this relatively large linewidth can be an advantage for biophotonic applications as it is spectrally narrow enough to address a given chromophore, but lacks the temporal coherence that would generate unwanted speckle patterns. *In fine* a continuous tunability is demonstrated between 440 and 670 nm using only five dye capsules (Fig.2 and 3).

Figure 4 shows the output power characteristics with a 10%-transmittance-output coupler in the case of Coumarin 540 A. Similar results were obtained with other materials. The small-signal absorption of the active layer was measured to be 75% in a single pass. The thresholds and slope efficiencies are given here with respect to the absorbed energy. A clear lasing threshold is visible at 11 µJ (8.46 mJ/cm$^2$) with maximum output energy of 4.5 µJ corresponding to a slope efficiency of 8%. This efficiency was identical in a VECSOL configuration (*i.e.* with the gain medium directly spin-coated onto the rear mirror) with the same cavity length, proving that the Brewster-angled gain chip and the tuning etalon only brought negligible additional losses. The modest value of the laser slope efficiency and the relatively high threshold are mostly attributed here to the long (15 mm) cavity. Importantly, the laser keeps its diffraction-limited output, which is one of the key advantages of VECSOLs in general over organic edge-emitting lasers. A typical image of the $TEM_{00}$ beam profile measured at the laser output is shown in the insert of Fig.4.

Finally, photostability was investigated at a fixed absorbed pump energy above threshold (20 µJ). The output emission decreased to half its initial value after $\approx 600$ pulses for a fixed position of the pump spot. The modest capsule lifetime is the result of the energetic UV pumping and the absence of any encapsulation. To maintain stable lasing performance during hours of operation, the gain capsule could be translated to illuminate a fresh spot after degradation. As an example, a 100-µm-in-diameter pump spot and a 25 mm×8 mm capsule allows ≈ 25 000 spots equivalent to ≈ 833 hours of operation at 10 Hz (≈ $3\times10^7$ pulses). Intrinsically the system is almost insensitive to misalignment: the gain medium is not bound to the resonator and is located at the pump laser waist and very close to the cavity beam waist, meaning that even though the gain capsule adds some phase distortion (because of *e.g.* surface inhomogeneity, possible thermal lensing, etc.) it will have only minor effect on the cavity alignment and laser pointing stability. Therefore, translation or replacement of the capsule has virtually negligible consequences on laser action, if adequate mechanical mounts and reasonably flat glass substrates are used. The low sensitivity can also be explained by the large gain of each capsule, a feature already shown in VECSOLs[28], where lasing was demonstrated with output couplers of reflectivities lower than 80%. In our experiment, the replacement of a chip by another was usually not associated with a loss of lasing. However, we observed that a continuous scanning of the sample led to some regions where a slight realignment was necessary, certainly because of the lack of parallelism of our very basic microscope slide over large areas.

This type of laser structure can consequently be thought of as an innovative solution to photodegradation issues if the gain capsule is regarded as a disposable element. Adding the costs of dye powder, optical grade PMMA solution and glass slide, the cost of a single capsule does not exceed 1 euro. In this context it is useful to consider that most efforts directed towards solving photodegradation issues have been focused either on the development of encapsulation techniques[24] or on improved chemical designs of the chromophores. However, enhancing stability and lifetime at any price might be based on an inadequate idea of the targeted applications for OSSLs. Indeed, as the latter will never be able to compete with crystalline or semiconductor solid-state systems in terms of reliability and stability, most applications that strongly demand a truly tunable visible laser (*e.g.* biophotonics, spectroscopy, sensing, or more generally applications based on image capture or measurement), will need a laser with a stable intensity only for a limited amount of time, that is maybe only a few thousands pulses, depending of course on the required signal-to-noise ratio, which is absolutely compatible with OSSLs operating in ambient conditions such as the one presented in this paper.

In conclusion, we demonstrated a new concept of UV-pumped organic thin-film laser widely tunable between 440 and 670 nm with a diffraction-limited output, using an intracavity Fabry-Perot etalon for fine tuning and an extremely simple and low-cost gain medium made of a microscope slide covered with a dye-doped PMMA film. The open cavity allows the emission range to be further extended to UV range using an intracavity frequency doubling as demonstrated in [ref. 29]. Output energy in the µJ-range with an efficiency of 8% has been obtained. The originality of the concept stems on the use of disposable low-cost and easy-to-make gain capsules. Each capsule provides several hours of laser operation and can be replaced once completely degraded without realignment of the laser cavity. Moreover, the cost of the whole laser structure (including pump source) could be lowered significantly in the near future thanks to the advent of high-brightness UV-blue diode lasers. This kind of widely-tunable laser source is likely to find applications in spectroscopy or sensing, where a high-brightness laser beam at a controllable visible wavelength is needed, but where stability is not a stringent requirement beyond the time needed for a single set of measurements. Also, being a colorful laser, straightforward to build with simple elements, easy to align and very close to a "textbook" linear cavity laser, it is an interesting tool for educational use.

The authors wish to acknowledge the Réseau des Opticiens (C.N.R.S.) and the A.N.R ("Vecspresso" Emergence program) for funding this work, and Paris 13 University for funding Dr Mhibik's post-doctoral position.

**FIGURES:**

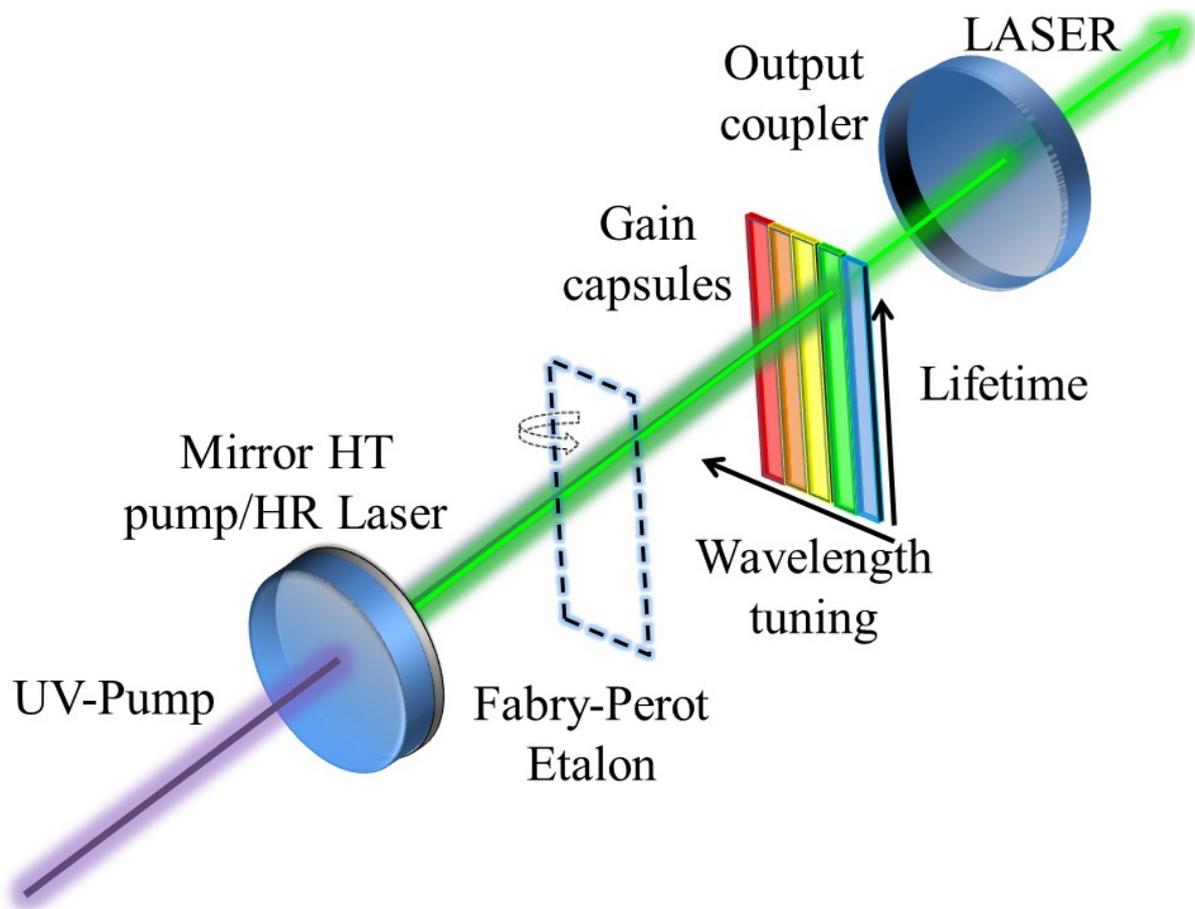

**Fig.1:** Laser architecture: Lateral translation of the capsule ensures access to different regions of the spectrum through the selection of given material, while rotating the Fabry-Perot intracavity etalon allows fine and continuous tunability within the 50-nm tuning range of each material. Vertical translation of the sample is a way to address different spots and to consequently increase the lifetime of the capsule.

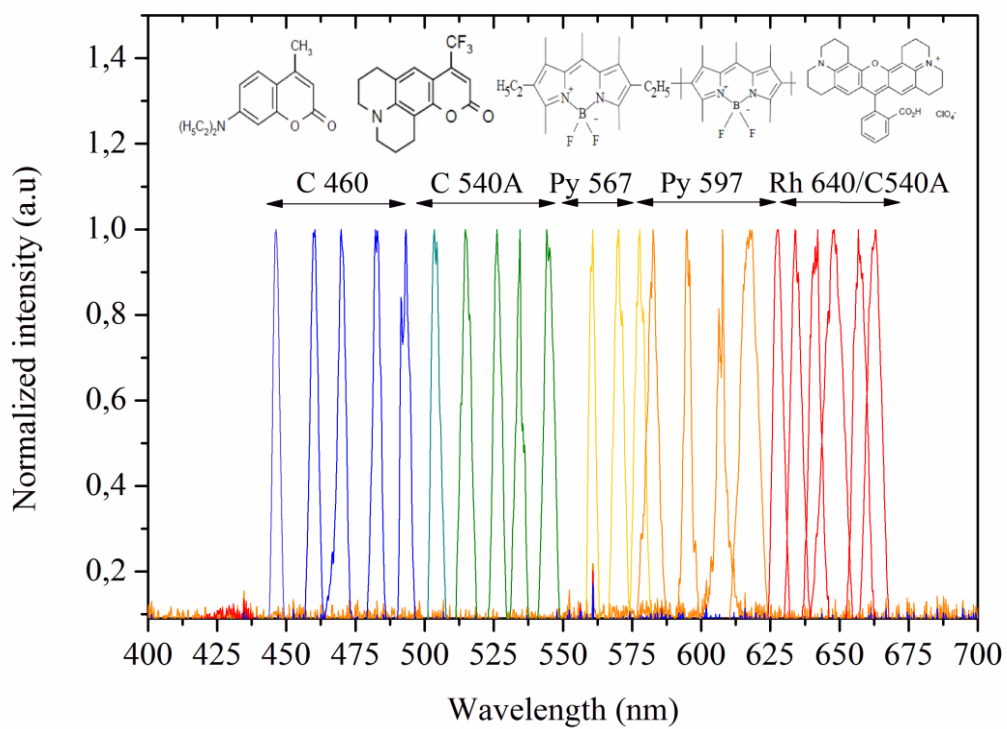

**Fig.2:** (color online) Tunability results with five capsules: coumarin 460 (C 460, blue solid line), coumarin 540 A (C 540A, green), pyrromethene 567 (Py 567, light orange solid line), pyrromethene 597 (Py 597, dark orange solid line), and with a blend of rhodamine 640 (Rh 640) and C 540A (red solid line). Continuous tunability is obtained through a slight tilting of 2μm-thick free-standing polymer film acting as an intracavity Fabry-Perot etalon.

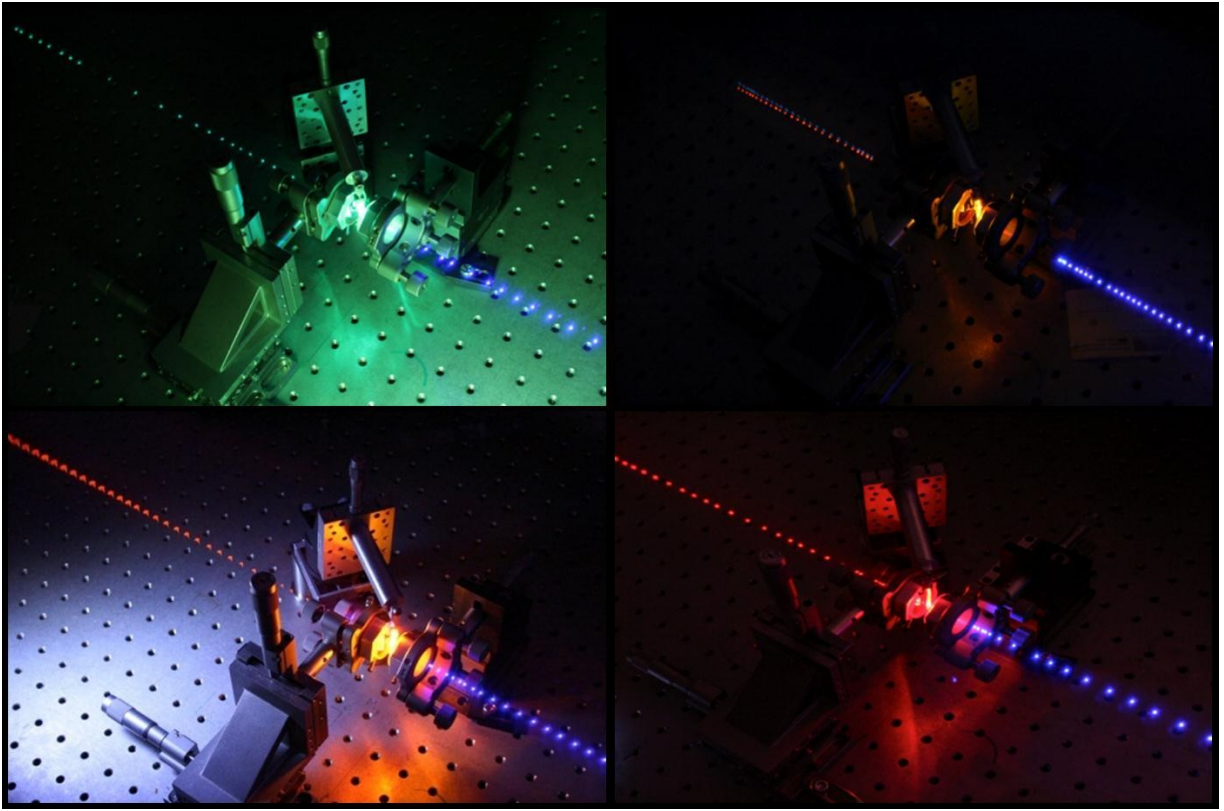

**Fig.3:** long-exposure photograph of the setup showing the pump beam (coming from the right, visible through fluorescence of a blank paper sheet), and the output coupled beam (on the left): from upper left and turning clockwise, emitting materials under use are coumarin 540 pyrromethene 567, Rhodamine 640, and pyrromethene 597

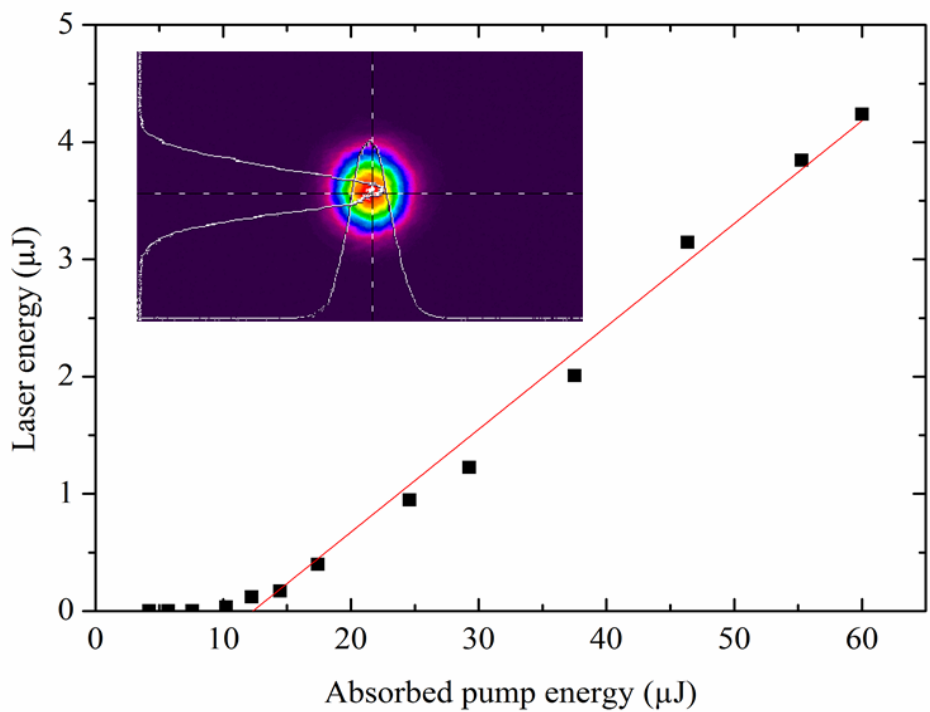

**Fig.4:** Output laser energy versus absorbed pump energy for the Coumarin 540A laser (Cavity length 16 mm, output coupler R=90%). The red line is a linear fit to the experimental data. Insert: Diffraction limited beam profile measured at the exit of the laser by a CCD camera.

**REFERENCES**

[1] T.Woggon, S. Klinkhammer and U. Lemmer, Appl. Phys. B 99, 47 (2010).

[2] D. Schneider, T. Rabe, T. Riedl, T. Dobbertin, M. Kröger, E. Becker, H.-H. Johannes, W. Kowalsky, T. Weimann, J. Wang, P. Hinze, A. Gerhard, P. Stössel, H. Vestweber, Adv. Mater. 17, 31 (2005).

[3] T.Woggon, S. Klinkhammer, and U. Lemmer, Appl. Phys. B 99, 47 (2010).

[4] C. Ge, L. Meng, W. Zhang, and B. T. Cunningham, Appl. Phys. Lett. 96, 163702 (2010).

[5] Y. Yang, G.A. Turnbull, and I.D.W. Samuel, Adv. Funct. Mater. 20, 2093 (2010).

[6] A. Rose, Z. G. Zhu, C. F. Madigan *et al.* Nature 434, 876 (2005).

[7] O. Mhibik, T-H My, D. Pabœuf, F. Bretenaker, and C. Drag, Opt. Lett. 35,2364 (2010).

[8] S. Forget, F. Balembois, G. Lucas-Leclin, F. Druon and P. Georges, Opt. Comm. 220, 187 (2003).

[9] I. Breunig, D. Haertle and K. Buse, Appl. Phys. B 105, 99 (2011).

[10] S. Chénais and S. Forget, Polym. Int. 61, 390 (2012).

[11] I. D. W. Samuel and G. A. Turnbull, Chem. Rev. 107, 1272 (2007).

[12] R. Bornemann, E. Thiel, and P. H. Bolívar, Opt. Express 19, 26387 (2011).

[13] M.Faloss M. Canva, P. Georges, A. Brun, F. Chaput, and J.-P. Boilot, Appl. Opt. 36, 6760 (1997).

[14] A. Costela, I. Garcia-Moreno, M. Figuera, F. Amat-Guerri, and R. Sastre, Laser Chem. 18, 63 (1998)

[15] I. Garcia-Moreno, A. Costela, V. Martin, M. Pintado-Sierra, and R. Sastre, Adv. Funct. Mater. 19, 2547 (2009).

[16] S. Klinkhammer, N. Heussner, K. Huska, T. Bocksrocker, F. Geislhöringer, C. Vannahme, T. Mappes, and U. Lemmer, Appl. Phys. Lett. 99, 023307 (2011).

[17] B. H. Wallikewitz, G. O. Nikiforov, H. Sirringhaus, and R. H. Friend, Appl. Phys. Lett. 100, 173301 (2012).

[18] S.Klinkhammer, T. Grossmann, K. Lüll, M.Hauser, C. Vannahme, T.Mappes, H. Kalt, U.Lemmer, IEEE Photonics Tech. Lett. 23, 489 (2011).

[19] M. Stroisch, T. Woggon, C. Teiwes-Morin, S. Klinkhammer, K. Forberich, A. Gombert, M. Gerken, U. Lemmer, Opt. Express 18, 5890 (2010).

[20] V. Navarro-Fuster *et al.* J. Appl. Phys. 112, 043104 (2012)

[21] B. Wenger, N. Tétreault, M. E. Welland, and R. H. Friend, Appl. Phys. Lett. 97, 193303 (2010).

[22] P.Görrn, M.Lehnhardt, W.Kowalsky, T.Riedl and S.Wagner, Adv. Mater. 23, 869 (2011).

[23] G. Heliotis, R. Xia, D. D. C. Bradley, G. A. Turnbull, I. D. W. Samuel, P. Andrew, and W. L. Barnes, Appl. Phys. Lett. 83, 2118 (2003).

[24] D. Schneider *et al.* Appl. Phys. Lett. 84, 4693 (2004).

[25] P. E. Burrows, V. Bulovic, S. R. Forrest, L. S. Sapochak, D. M. McCarty, and M. E. Thompson, Appl. Phys. Lett. 65 (1994).

[26] H. Rabbani-Haghighi, S. Forget, S. Chénais, and A. Siove, Opt. Lett. 35, 1968 (2010).

[27] M.Berggren A. Dodabalapur, and R. E. Slusher, Appl. Phys. Lett. 71, 2230 (1997).

[28] H. Rabbani-haghighi, S. Forget, A. Siove and S. Chénais, Eur. Phys. J. Appl. Phys. 56, 34108 (2011).

[29] S. Forget, H. Rabbani-Haghighi, N. Diffalah, A. Siove, and S. Chénais, Appl. Phys. Lett. 98, 131102 (2011).

---

[a] Author to whom correspondence should be addressed. Electronic mail: sebastien.forget@univ-paris13.fr.